# Rotational excitations in concentrated solid Kr–CH₄ solutions. Calorimetric studies

M.I. Bagatskii, V.G. Manzhelii, I.Ya. Minchina, D.A. Mashchenko, I.A. Gospodarev.

*Verkin Institute for Low Temperature Physics and Engineering, National Academy of Sciences of Ukraine, 47, Lenin Ave., 61164, Kharkov, Ukraine.*

*Email: bagatskii@ilt.kharkov.ua*

*The heat capacity of solid Kr–CH₄ solutions with 30 and 60 mol.% CH₄ has been studied at 0.8–20 K. The contribution of the rotational subsystem $C_{rot}$ to the heat capacity of the solutions is separated. The results obtained in this study and Ref. [4] were used to estimate the difference between the lowest-level energies $\varepsilon_{AT}$ of the nuclear spin A and T modifications of CH₄ and to find the characteristic conversion times $\tau$ for the solutions with 5–60 mol.% CH₄ at low temperatures.*

## 1. INTRODUCTION

The low temperature dynamics of the quantum rotors (CH₄) subsystem in concentrated solid $(CH_4)_n Kr_{1-n}$ solutions is investigated by the calorimetric method. Among inert gases krypton has the Lennard–Jones potential whose parameters approach most closely those of CH₄ [1,2]. CH₄ and Kr therefore form a continuous series of solid solutions at $n < 80$ mol.% CH₄ [2] (the segment of the CH₄–Kr equilibrium diagram in Fig. 1 [3]). This allows investigations not only of isolated molecular rotors but of their interactions as well. The dependence of the translation oscillations of the solution on CH₄ concentration n can be taken into account in the quasi-isotropic approximation.

Solid Kr–CH₄ solutions have been studied quite intensively. The studies of the thermophysical properties of the Kr–CH₄ solutions are surveyed in the Reference Handbook [2]. Recently, the main attention has been concentrated on the rotational spectra and the rate of the nuclear-spin conversion of $CH_4$ molecules in these solutions (see below). However, the 6-60 mol.% $CH_4$ concentration region lacked close approach, which impeded estimation of the rotor interaction effect upon the solution dynamics. The main goal of this study is to obtain information about the characteristic conversion times $\tau$ and the low-energy part of the rotational spectrum of the rotor subsystem in concentrated Kr-$CH_4$ solutions at liquid helium temperatures. The problem and the task are detailed below.

Earlier, the heat capacity of solid Kr–$CH_4$ solutions was studied in low concentrated solutions with $1, 5, 10$ mol.% $CH_4$ and in the $(CH_4)_{0.01}Kr_{0.988}(O_2)_{0.002}$ solution at $0.7$–$8$ K [4]. We found and analyzed qualitatively the correlation between the experimentally detectable heat capacity and the conversion rate among three nuclear spin modifications of $CH_4$ (A, T, E) whose molecules have the total nuclear spins $I = 2, 1, 0$, respectively.

The qualitative regularities of nuclear spin conversion were found for non-concentrated solid Kr–$CH_4$ solutions. It was revealed that at low temperatures conversion is effected mainly by a hybrid mechanism [5]. According to Ref. [5] "the most effective intrinsic mechanism for free rotor molecules is the hybrid process in which the intramolecular dipole-dipole interaction mixes the nuclear spin states and the intermolecular octupole-octupole interaction causes transitions between rotational states and conserves energy by coupling to the lattice". This is manifested by the fast growth of the conversion rate when the $CH_4$ concentration and hence the number and size of the $CH_4$ clusters increase in the solution.

Even small amounts of paramagnetic $O_2$ admixture accelerate the conversion significantly.

Below 3 K the conversion is determined by on the transitions between the ground A- and T- states (see below). The spacing $\varepsilon_{AT}$ between ground energy states of the A- and T- modifications of $CH_4$ in the Kr lattice were calculated fairly accurately from the calorimetric data for the solid $(CH_4)_{0.01}Kr_{0.988}(O_2)_{0.002}$ solution.

No orientational glass with an indirect interaction was detected in non-concentrated solid $CH_4$–$Kr$ solutions [4]. In such glasses the rotational heat capacity of the impurity subsystem is linearly dependent on the temperature and independent of the impurity concentration [6].

The previous investigation [4] is extended here to include higher $CH_4$ concentrations (30%, 60%) and a wider temperature interval (0,8–20 K). Doing so we expected in particular to use the results obtained along with Ref. [4] and literature data for calculating the total concentration dependence $(0 < n < 1)$ of the conversion rate and for evaluating $\varepsilon_{AT}$ for the $(CH_4)_nKr_{1-n}$ solutions at low temperatures. Another task was to search for orientation glass in $CH_4$–$Kr$ solutions.

It should be noted that concentrated $(CH_4)_nKr_{1-n}$ solutions have enjoyed much less experimental attention that their counterparts with low $CH_4$ and $Kr$ concentrations (the vicinity of the $T$–$n$ phase diagram edges in Fig. 1 [3]). Besides, it is easier to interpret the results for weak solutions since the rotational energy spectra of the molecules in the solid methane and isolated $CH_4$ molecules in the solid $Kr$ matrix have been calculated theoretically [7,8]. There is also experimental information about the $CH_4$ spectra for these limiting cases [8–13].

The rotational energy spectra of the A-, T-, E- modifications in the octahedral crystal field are shown in Fig. 2 [8]. The A- modification has lowest ground energy state; as a result, the $CH_4$ molecules all appear to be in this state at $T = 0$ K. At least down to 3 K $C_{rot}$ is determined only by the transitions between the lowest levels of the A- and T- modifications, the change in the energy being $\varepsilon_{AT} = \varepsilon_{0T} - \varepsilon_{0A}$.

Several factors are responsible for the dynamics of rotors in concentrated $CH_4$–$Kr$ solutions and this complicates the interpretation of calorimetric results. Concentrated solutions contain many variants of clusters ranging in size and arrangement of $CH_4$ molecules. The energy spectra of $CH_4$ molecules therefore vary widely because $CH_4$ molecules are exposed to different local symmetries and their non-central interactions with the surroundings are different too. It is very difficult to allow for these non-central interactions even if they are reduced to the nearest neighbors. Because of frustration, the non-central interaction is dependent not only on the number of nearest neighbors, but on their mutual arrangement as well. Besides, it is important that the octupole moment of the $CH_4$ molecule is dependent on its energy level. In the rotational ground state, the $CH_4$ molecule has the minimal octupole moment. It increases when the $CH_4$ molecules change into excited states [7]. As the temperature of the Kr–$CH_4$ solution rises, the occupancy of the excited states of the $CH_4$ rotors increases. So do the octupole moment and the interaction of $CH_4$ molecules, i.e. the $CH_4$ interaction changes with temperature.

Our calorimetric investigation was expected to show how the above factors influenced the conversion and the low-energy part of the $CH_4$ rotation spectrum in concentrated $CH_4$–Kr solutions.

## 2. EXPERIMENT

The heat capacity of solid $CH_4$–Kr solutions with the $CH_4$ concentrations $n = 29.7$ mol.% and $59.9$ mol.% was studied at $0.8$–$20$ K by pulse heating using an adiabatic vacuum calorimeter [15]. The change in the sample temperature $\Delta T$ during one measurement of heat capacity was about 10% of the initial $T_i$. The heating time $t_h$ was 2–6 min. The effective time $t_m$ of one heat capacity measurement was $t_m = t_h + t_e$, where $t_e$ is the time taken to achieve a steady time dependence of temperature operation of the calorimeter since the moment of switching off the heating. The sample masses ($0.4725$ mole ($n = 29.7\%$) and $0.4412$ mole ($n = 59.9\%$)) and the concentrations were found from the $PVT$ data for the gaseous components, the error being 0.2%. The gas purity was: $CH_4$ – 99.94% (0.04% $N_2$, $\leq 0.01\%$ $O_2$, and Ar), Kr – 99.72% (0.08 $N_2$, 0.2% Xe,

< 0.01% $O_2$ and Ar). The solid solutions were prepared in a calorimeter at $T \approx 70$ K by condensing the gas mixture to the solid phase. This technology ensured homogeneous solutions. The measurement error in the heat capacity was 6% at 0.8 K, 2% at 1 K, 1% at 2 K and 0.5% above 4 K.

In this paper the term "heat capacity" is used for the derivative of the heat transmitted to the system with respect to temperature, no matter whether the system is equilibrium or not.

The heat capacity of phase I (see Fig. 1) of a solid $CH_4$–Kr solution was measured by A. Eucken and H. Veight [16] in 1936 ($n$ = 28 mol.%, $T$ = 12–25 K). Our results differ from theirs by 10%, which may by due to possible distinctions in the used temperature scales and deviations from equilibrium.

The rotational heat capacity $C_{rot}$ was obtained by subtracting the translational lattice component $C_{tr}$ from the heat capacity $C_{sol}$ of the solution. In the 30% $CH_4$ case $C_{tr}$ was assumed to be $C_{tr} = C_{tr,Kr} + \Delta C_{tr,Kr}$, where $C_{tr,Kr}$ is the heat capacity of pure Kr, $\Delta C_{tr,Kr}$ is the change in the translational heat capacity due to the introduction of a lighter $CH_4$ impurity into the Kr lattice. For the solution with 60% of $CH_4$ we assumed $C_{tr} = C_{tr,CH4} + \Delta C_{tr,CH4}$, where $C_{tr,CH4}$ is the translational component of the heat capacity of pure methane, $\Delta C_{tr,CH4}$ is the change translational heat capacity after the heavy Kr impurity was introduced into the $CH_4$ lattice (the contribution of quasi-local oscillations). $C_{tr,CH4}$ was calculated on the basis of the Jacobian matrix properties [17] and the characteristic temperature $\Theta$ = 140 K. $\Delta C_{tr,Kr}$ and $\Delta C_{tr,CH4}$ was calculated by the Jacobian matrix method [18] without regard for the changes in the force constants for the mass ratios $m_{CH4}/m_{Kr} = 0.2$ and $m_{Kr}/m_{CH4} = 5$ for 30% $CH_4$ and 60% $CH_4$ respectively.

The behavior of $C_{rot}(T,n)$ is analyzed comprehensively for temperatures below 3 K. At $T$ = 2 K the contribution of $C_{rot}(T,n)$ to $C_{sol}$ is 12% for $n$ = 5%, 61% for $n$ = 10%, 96% for $n$ = 30% and 99% for $n$ = 60%. The heat capacity of pure Kr ($C_{Kr}$ ) was measured by us previously at 0.7 – 20 K. The $C_{Kr}$ results agree well with the data of [41,42]. The limiting Debye temperature at $T \to 0$ is $\Theta_0 = 71.6$ K and coincides, within the experimental error, with $\Theta_0 = 71.7$ K [41] and $\Theta_0 = 71.9$ K [42].

## 3. RESULTS AND DISCUSSION

The experimental $C_{sol}$ values for equilibrium vapor elasticity of solid Kr-$CH_4$ solutions with $n = 4.93$, $9.82$ [4,43]; $29.7$ and $59.9$ mol.% $CH_4$ (below $n = 5$, $10$, $30$ and $60\%$) are given in Table 1.

The heat capacities $C_{rot}$ of the solution with $n = 4.93$, $9.82$ [4], $29.7$ and $59.9$ mol.% $CH_4$ normalized to $n$ and the universal gas constant $R$ are shown in Fig. 3. As the temperature rises, the $CH_4$ rotation approaches the free rotation and the conversion rate increases rapidly [19]. As a result, the "high temperature" heat capacity is close to the equilibrium and $C_{rot}$ should tend to the limiting value for the free rotor ensemble *3R/2* (see Fig. 3).

It is found experimentally [4] that the contribution of isolated $CH_4$ molecules (singles) to $C_{rot}$ of the solid Kr–$CH_4$ solution is negligible at liquid helium temperatures. The explanation is as follows. At these temperatures the characteristic conversion time $\tau$ of isolated $CH_4$ molecules is much longer then the effective time of one heat capacity measurement $t_m$. Before the experiment, the sample was kept at lowest temperature for about 24 hours so that most $CH_4$ molecules could convert to the ground state of the A- modification. Since the spacing between the ground and first excited states is large in this modification ($\sim 55$ K [8], see Fig. 2), the overwhelming majority of singles remain in the ground state at liquid helium temperature.

The conversion rate in $CH_4$ clusters is controlled by the aforementioned hybrid mechanism [5] and proceeds much faster than that of singles [4], as a result the contribution of the clusters to $C_{rot}$ becomes dominant. The heat capacities $C_{rot}$ normalized to the gas constant $R$ and the $CH_4$ concentration $n' = n(1-(1-n)^{12})$ in the clusters are shown in Fig. 4 (on an enlarged scale down to $T = 4.3$ K). The $C_{rot}/Rn$ (Fig. 3) and $C_{rot}/Rn(1-(1-n)^{12})$ (Fig. 4) values practically coincide for concentrated solutions since isolated $CH_4$ molecules are few in these solutions (see Table 2).

The behavior of $C_{rot}$ varies in solutions with different $CH_4$ concentrations depending on the particular conversion rate and the energy spectra of $CH_4$ molecules in each solution. This offers the possibility of obtaining information about the energy spectrum and characteristic conversion times from the heat capacity data.

So far there are no theoretical calculation for the energy spectra of $CH_4$ rotors in clusters (even two-molecule ones). The $CH_4$ rotation in Kr–$CH_4$ solutions at low temperatures was studied by different methods – inelastic neutron scattering (INS) [9,10,19-22], nuclear magnetic resonance (NMR) [11,12,13], spectroscopic methods [24,25]. The INS and NMR data were discussed in the one-particle approximation within a simple mean potential field model for a random distribution of the solution components (Kr,$CH_4$) over the lattice sites.

At $T < 3$ K, $C_{rot}(T,n)$ is determined only by the transitions of the $CH_4$ molecules between the lowest levels of the A- and T- modifications (see Fig. 2), the change in the energy being $\varepsilon_{AT} = \varepsilon_{0T} - \varepsilon_{0A}$. We can therefore use the expression for the heat capacity of a two-level system.

$$C_{rot}(T,n) = N\,k_B(\varepsilon_{AT}/T)^2(g_1/g_0)exp(-\varepsilon_{AT}/T)/(1 + (g_1/g_0)\,exp(-\varepsilon_{AT}/T))^2, \quad (1)$$

where $N$ is the number of $CH_4$ molecules in the sample ($N = N_0 n$ per mole, where $N_0$ is Avogadro number), $k_B$ is the Boltzmann constant, $g_0$ and $g_1$ are the degeneracies of the ground and the first excited levels, respectively.

$C_{rot}$ measured for Kr–$CH_4$ solutions at $T \leq 2.5$ K is shown in Fig. 5 in the $ln(C_{rot}T^2) \div 1/T$ coordinates (the denominator in Eq. (1) is negligible at $T \leq 2.5$ K and $\varepsilon_{AT} \approx 10$ K). The linear dependence $ln(C_{rot}T^2) \div 1/T$ holds, within the experimental error, for all solutions with $n$ = 4.93, 9.82, 29.7 and 59.9 mol.% $CH_4$. The effective $\varepsilon_{AT}$ was found from the straight line slopes (see Table 3). The $\varepsilon_{AT}(n)$ values ([4] and these study) are shown in Fig. 6 along with INS [9,10,13,14,20,26] and NMR [11,12] data for temperatures below 5 K. The data [11,12] for $n$ = 60–85% $CH_4$ solutions are rather inaccurate since they were obtained from the best-description condition for the temperature dependence of nuclear magnetic susceptibility in a very wide interval (0.5-20 K). At low $CH_4$ concentrations in krypton, the calorimetric and INS [9,10] values of $\varepsilon_{AT}$ agree well within the measurement error. For the 60% $CH_4$ solution, $\varepsilon_{AT}$ corresponds closely with $\varepsilon_{AT}$ extrapolated from the dependence $\varepsilon_{AT}(n)$ based on the data of [10,13,14,20,26].

As the number of $CH_4$ molecules changes in the cluster, so do the local symmetry and the crystal and molecular mean potential fields. The presence of different clusters in the Kr–$CH_4$ solutions should lead to splitting and smearing of the lines in the energy spectrum of $CH_4$ rotors. We however see (Fig. 5) that Eq. (1) is obeyed well, which suggests that below 3 K the splitting and smearing of the ground state are quite modest for the T-modification even in the 30% $CH_4$ and 60% $CH_4$ solutions. This also shows that the temperature dependence of the characteristic conversion time $\tau$ is negligible at 0.8–3 K up to $n$ = 60% $CH_4$.

The dependence $\varepsilon_{AT}(n)$ is mainly controlled by the molecular mean field, which changes when Kr atoms are replaced by $CH_4$ molecules. The nonmonotone dependence (Fig. 6) of the effective $\varepsilon_{AT}$ on the $CH_4$ concentration in the low-temperature part of disordered phase I (Fig. 1) is the manifestation of the nonmonotone concentration dependence of the molecular field. At low $n$ the increase in the number of $CH_4$ neighbors around a $CH_4$ molecules leads to an increase in the non-central forces acting on this molecules. As a result, $\varepsilon_{AT}$ decreases at growing $n$. However, at high $n$ the non-central forces induced by the surrounding molecules can be partially or completely counterbalanced [7,14,20]. At high $n$ ($n$ >0.5) the effect increases with the $CH_4$ concentration. $\varepsilon_{AT}$ grows approaching $\varepsilon_{AT}$ for a free rotor. The effect is particularly pronounced in the low temperature phase II of solid $CH_4$ in which one fourth of the molecules behave much like free rotors. The other three fourths are orientationally ordered. The existence of two basically distinct types of sublattices accounts for two an-order-of-magnitude different $\varepsilon_{AT}$-values both in pure $CH_4$ and in the adjacent phase II of the solid Kr – $CH_4$ solution (see Figs. 1, 6).

Let us estimate how much the experimental heat capacities $C_{rot}^{exp}$ deviate from the equilibrium $C_{rot}^{theor}$ in the low temperature region ($T < 3$ K) where Eq. (1) holds. In this region $C_{rot}^{exp}$ is determined by the change in the ground state occupancy of the T- modification during the effective time $t_m$ of one heat capacity measurement.

The $C_{rot}^{exp} / C_{rot}^{theor}$ ratios are shown in Fig. 7. $C_{rot}^{theor}$ was calculated for $\varepsilon_{AT}$ = 11.7 K ($n = 5\%$), 10.8 K ($n = 10\%$), 8.5 K ($n = 30\%$), 7.4 K ($n = 60\%$) by Eq. (1). $K' = C_{rot}^{exp} / C_{rot}^{theor}$ equals the ratio of the number of $CH_4$ molecules which underwent the $\varepsilon_{0A} \rightarrow \varepsilon_{0T}$ transformation during the time $t_m$ in the real experiment to the corresponding number of molecules for the equilibrium A-T-E distribution.

At liquid helium temperatures $C_{rot}^{exp}$ is determined only by the conversion of the $CH_4$ molecules in clusters (see above). The matrix-isolated molecules contribute negligibly [4]. In this case it is reasonable to use $K'' = K'/(1-(1-n)^{12})$ which is similar to $K'$ when only clustered molecules are considered. At high concentrations of methane $K'$ and $K''$ practically coincide (see Table 3).

The information obtained can be used to estimate the characteristic conversion time $\tau$ in clusters at $T < 3$ K. Note that before experiment the sample was kept for about 24 hours at the lowest measurement temperature so that the overwhelming majority of $CH_4$ molecules could convert into the ground state of the A-modification. Relaxation of the rotor subsystem in the clusters to the equilibrium distribution of the nuclear-spin modifications is described as [19,27]

$$N(T_{i+1}, \infty) - N(T, t) = (N(T_{i+1}, \infty) - N(T_i, 0))exp(-t/\tau), \qquad (2)$$

where $\tau$ is the characteristic time of A – T ground state conversion, $T_i$ is the sample temperature at the moment of switching on the heating ($t$=0), $T_{i+1}$ is the heating – produced temperature. The time needed to achieve $T_{i+1}$ is much longer than $\tau$. In Eq.(2) it is $t = \infty$. $N(T_i,0)$ and $N(T_{i+1},\infty)$ mean the equilibrium number of the $CH_4$ molecules with the energy $\mathcal{E}_{0A}$ in the clusters at $T_i$ and $T_{i+1}$. $N(T,t)$ is the number of the $CH_4$ molecules with the energy $\mathcal{E}_{0A}$ in the clusters at $T_i < T < T_{i+1}$ at the instant of time t. We write Eq. (2) for the instant of time $t = t_m$. In the left-hand side of Eq. (2) we add and subtract $N(T_i,0)$:

$$[N(T_{i+1},\infty) - N(T_i,0)] + [N(T_i,0) - N(T,t_m)] = [N(T_{i+1},\infty) - N(T_i,0)] \times exp(-t_m/\tau).$$

This expression can be re-written as:

$$1 - \{[N(T,t_m) - N(T_i,0)]/[N(T_{i+1},\infty) - N(T_i,0)]\} = exp(-t_m/\tau). \qquad (3)$$

Since we describe the conversion of $CH_4$ molecules in clusters, the second term in the left-hand side of Eq.(3) is $K''$. This relation can be re-written as:

$$1 - K'' = exp(-t_m/\tau).$$

Which gives:

$$\tau = -t_m/ln(1-K'') \qquad (4)$$

The effective times $\tau$ calculated by Eq. (4) for the solutions with $n = 5$, 10, 30, 60 mol.% $CH_4$ are summarized in Table 3 and in Fig. 8 (along with literature data [11,19,29,30]). The calculation by Ref. [4] for $(CH_4)_{0.01}Kr_{0.988}(O_2)_{0.002}$ is shown too (the paramagnetic $O_2$ admixture stimulates conversion). But in the latter case Eq. (4) should have $K''$ instead of $K'$.

Let us remember that Eq.(2) is written using a unified characteristic time $\tau$ to describe the conversion of $CH_4$ molecules in $CH_4$ clusters varying in size and shape. This implies that $\tau$-values calculated by Eq. (4) are effective (averaged over the clusters) times of conversion for the solutions investigated.

The $\tau$ value that we measured for the $n = 5\%$ $CH_4$ solution is about four times lower than the INS- based result for the $n = 0.3\%$ $CH_4$ case [19]. This is not in conflict with the data of [31,32] and supports [4,33] our conclusion about predominance of the hybrid mechanism of conversion at low temperatures [5]. In the solution with $n = 0.3\%$ $CH_4$ nearly all $CH_4$ molecules are matrix-isolated. According to [5], their conversion is much slower than that of clustered $CH_4$ molecules.

It is seen in Table 3 that the $O_2$ admixture has accelerated considerably the conversion of isolated $CH_4$ molecules in the Kr matrix, which agrees qualitatively with theoretical [5] and experimental [4,27,33-39] data for the $O_2$ admixture effect upon the conversion in solid $CH_4$ and Kr–$CH_4$ solutions.

The conversion of isolated $CH_4$ molecules in the Kr matrix can be stimulated by the isotope $^{83}$Kr, too (11.56% in natural gas, the nuclear spin is 9/2 atomic units). Earlier, the effect of matrix isotopes with non-zero nuclear spins upon the conversion of the molecules introduced into solid inert gases was approached theoretically only for linear $^{14}N_2$ and $^{15}N_2$ molecules [40]. We did not detect any $^{83}$Kr effect during the time $t_m$ [4], which implies that $^{83}$Kr is a much weaker conversion-stimulator than the hybrid mechanism or the paramagnetic $O_2$ admixture.

As to the orientational octupole glass, no clear evidence of its presence was found in the solutions with $n \leq 60\%$ $CH_4$ (see Figs. 3, 4). This suggests that below 3K the splitting and smearing of the lower energy levels of quantum $CH_4$ rotors caused by their octupole-octupole interaction and the local symmetry violations are smaller than the mean spacing $\varepsilon_{AT}$ of these spectral levels.

It is therefore interesting to continue a search for octupole glass at low temperatures and $CH_4$ concentrations near the phase II boundary.

## CONCLUSIONS

Solid $CH_4$–Kr solutions enable us to study the dynamics of quantum rotors ($CH_4$) in the crystal in a wide range of temperatures and rotor concentrations. The quantum effects of the rotor subsystem are most pronounced at liquid helium temperatures. Of special interest are the energy spectra of the rotational motion and conversion in the rotor subsystem. This information permits us to estimate the thermodynamic properties of the rotor subsystem and to find the time necessary to achieve the equilibrium in the solution. Until recently, such information was unavailable for the concentration region $n = 6 - 60 \%$ $CH_4$. The goal of this study was to derive the lacking information at liquid helium temperatures and to discuss the dependences obtained.

Proceeding from the heat capacity data for solid $CH_4$–Kr solutions measured in this study and in [4], we were able to separate the contributions of the rotor subsystem to the heat capacity. Based on the analysis of the contributions, the concentration dependences were obtained for the characteristic conversion time $\tau(n)$ and the energy difference $\varepsilon_{AT}(n)$ between the rotational ground states of the A– and T– modifications of $CH_4$. The value $\varepsilon_{AT}$ just determines the contribution of the rotor subsystem to the thermodynamic properties of the solution at liquid helium temperatures.

The dependences $\varepsilon_{AT}(n)$ and $\tau(n)$, along with literature data for $n < 6 \%$ and $n > 60\%$ $CH_4$, cover the whole concentration region where solid $CH_4$–Kr solutions can exist. The non-monotonic dependence $\varepsilon(n)$ in orientationally disordered phase I can be attributed to the non-monotonic dependence of the molecular field on the $CH_4$ concentration $n$. At low $n$ the increasing number of $CH_4$ neighbors around a $CH_4$ molecule enhances the molecular field. $\varepsilon_{AT}$

decreases with growing $n$. At high $n$, however, because of frustration, the effects counterbalancing the non-central forces of the $CH_4$ neighbors upon the surrounded $CH_4$ molecule become strong [7,14,20]. As a result, the molecular field is depressed and $\varepsilon_{AT}$ increases approaching the value typical for a free rotor.

Note that at high $n$ ($n > 0.8$) there is also a mechanism enhancing the molecular field and much decreasing $\varepsilon_{AT}$ at growing $n$. This mechanism however operates in phase II of the solid solution in the sublattices with a long-range orientational order [7,13].

There is experimental evidence in [4] that the $CH_4$ conversion is much faster in clusters than in isolated $CH_4$ molecules (singles). This is in qualitative agreement with the theoretical predictions by Nijman and Berlinsky [5] about the predominant hybrid mechanism of conversion for weakly retarded $CH_4$ molecules. According to this mechanism, the increase in the non-central interaction of $CH_4$ molecules in the clusters leads to shorter $\tau$. As a result, the non-monotonic dependence $\tau(n)$ may be attributed to the above non-monotonic concentration dependence of the molecular field acting upon the $CH_4$ molecule.

The proposed qualitative interpretation of the dependences $\varepsilon_{AT}(n)$ and $\tau(n)$ cannot be taken as a rigorous consideration as it does not allow for some factors, such as the diversity of clusters, dependence of the molecular field value and symmetry upon the cluster shape and size. In any event the problem is very complicated, and we can hardly expect a rigorous theory for solid $CH_4$–$Kr$ solutions in the nearest future.

Two more results of this study are worthy of notice here. It is shown that the values of splitting and smearing of the ground state of the T– modification of $CH_4$ are small and not obvious in the behavior of heat capacity at helium temperatures, even in the solution with 60 % $CH_4$.

No octupole glass was detected in the solution with $n = 60$ % $CH_4$.

A further increase in the $CH_4$ concentration will make the splitting of the ground state of the T– modification inevitably obvious, and detection of octupole glass will be quite probable.

We are planning to continue the low temperature investigation of the solid $(CH_4)_n Kr_{1-n}$ solutions with the $CH_4$ concentrations $n = 0.6 – 1$.

## ACKNOWLEDGEMENTS

The authors are indebted to A.I. Prokhvatilov, M.A. Strzhemechny, and A.I. Krivchikov for helpful discussions. The work was supported by the Ukraine Minister of Education and Science (Project "New quantum and anharmonic effects in crystal solutions", No 02.07/00391-2001).

## REFERENCES

1. V.G. Manzhelii, A.I. Prokhvatilov, V.G. Gavrilko, A.P. Isakina, *Structure and Thermodynamic Properties of Cryocrystals,* (Handbook, Begell House, inc., New York, Wallingford, (UK), 1998).

2. V.G. Manzhelii, A.I. Prokhvatilov, I.Ya. Minchina, L.D. Yantsevich, *Handbook of Binary Solutions of Cryocrystals,* (Begell House, inc., New York, Wallingford, (UK), 1996).

3. S. Grondey, *Eingefrorene Orientierungs-und Rotationsanregungen in festen Mischungen von Methan und Krypton (Neutronenstreuexperiment),* (Als Manuskript gedruckt, Berichte der Kernforschungsanlage Julich – Nr. 2083, Institut fur Festkorperforschung, Jul – 2083, 1986).

4. I.Ya. Minchina, V.G. Manzhelii, M.I. Bagatskii, O.V. Sklyar, D.A. Mashchenko, and M.A. Pokhodenko, *Low Temp. Phys.* **27**, 568 (2001) [*Fiz. Nizk. Temp.* **27**, 773 (2001)].

5. A.J. Nijman, A.J. Berlinsky, *Canad. J. Phys.* **58**, 8, 1049 (1980).

6. M.I. Bagatskii, V.G. Manzhelii, M.A. Ivanov, P.I. Muromtsev, I.Ya. Minchina, *Sov. J. Low Temp. Phys.* **18**, 801 (1992), [*Fiz. Nizk. Temp.* **18**, 1142 (1992)].

7. T. Yamamoto, Y. Kataoka, K. Okada, *J. Chem. Phys.* **66**, 2701, (1977).

8. K. Nishiyama, T. Yamamoto, *J. Chem. Phys*. **58**, 3, 1001 (1973).

9. B. Asmussen, W. Press, M. Prager, and H. Blank, *J. Chem. Phys.* **98**, 158 (1993).

10. B. Asmussen, W. Press, M. Prager, and H. Blank, *J. Chem. Phys.* **97**, 1332 (1992).

11. P. Calvani and H. Glattli, *J. Chem. Phys.* **83**, 1822 (1985).

12. P. Calvani and H. Glattli, *Solid State Commun.* **50**, 169 (1984).

13. Werner Press, *Rotational Excitations in Disordered Molecular Solids,* Dynamics of Molecular Crystals, 615, (1987).

14. S. Grondey and M. Prager, W. Press, *J. Chem. Phys.* **86**, 6465 (1987).

15. M.I. Bagatskii, I.Ya. Minchina, V.G. Manzhelii, *Sov. J. Low Temp. Phys.* **10**, 542 (1984), [*Fiz. Nizk. Temp.* **10**, 1039 (1984)].

16. A. Eucken, H. Veith, *Z. Phys. Chem*. *B* **34**, 275 (1936).

17. V.I. Peresada, *Zh. Eksp. Teor. Fiz.* (in Russian), **53**, 605 (1967).

18. V.I. Peresada, V.P. Tolstoluzhskii, *Sov. J. Low Temp. Phys*. **3**, 378 (1977), [*Fiz. Nizk. Temp.* **3**, 788 (1977)].

19. S. Grieger, H. Friedrich, B. Asmussen, K. Guckelsberger, D. Nettling, W. Press, and R. Scherm, *Z. Phys. B., Condens. Matter.* **89**, 203 (1992).

20. S. Grondey, M. Prager, W. Press and A. Heidemann, *J. Chem. Phys.* **85**, 2204 (1986).

21. W. Press, *J. Chem. Phys.* **56**, 2597 (1972).

22. K.J. Lushington, *J. Chem. Phys.* **76**, 3843 (1982).

23. F. De Luca and B. Maraviglia, *Chem. Phys. Lett.* **101**, 300 (1983).

24. Ken'ishi Nishiyama, *J. Chem. Phys.* **56**, 5096 (1972).

25. A. Kabana, G.B. Savitsky, and D.F. Hornig, *J. Chem. Phys.* **39**, 2942 (1963).

26. Werner Press, *Single-Particle Rotations in Molecular Crystals,* Vol. **92** (Springer Tracts in Modern Physics, Springer-Verlag, Berlin,

Heidelberg, New York, 1981).

27. H.P. Hopkins, P.Z. Donoho, K.S. Pitzer, *J. Chem. Phys.* **47**, 864 (1967).

28. A.I. Prokhvatilov, A.P. Isakina, *Fiz. Nizk. Temp.* **10**, 1206 (1984), [*Sov. J. Low Temp. Phys.* **10**, 631 (1984)].

29. J.E. Piott, W.D. McCormik, *Can. J. Phys.* **54**, 1784 (1976).

30. J. Higinbotham, B.M. Wood, and R.F. Code, *Phys. Lett.,* **66A**, 237 (1978).

31. V.V. Dudkin, B.Ya. Gorodilov, A.I. Krivchikov, and V.G. Manzhelii, *Low Temp. Phys.* **26**, 762 (2000), [*Fiz. Nizk.Temp.* **26**, 1023 (2000)].

32. A.I. Krivchikov, B.Ya. Gorodilov, V.V. Dudkin, and V.G. Manzhelii, *Third International Conference on Cryocrystals and Quantum Crystals,* (Abstracts and Programme) Sklarska Poreba, Poland, 64 (2000).

33. I. Minchina, V. Manzhelii, M. Bagatskii, O. Sklyar, D. Mashchenko, *Third International Conference on Cryocrystals and Quantum Crystals,* (Abstracts and Programme) Sklarska Poreba, Poland, 74 (2000).

34. F.H. Frayer and G.E. Eving, *J. Chem. Phys.* **48**, 781 (1968).

35. H. Glattli, A. Sentz, M.K. Eisen, *Phys. Rev. Lett.* **28**, 871 (1972).

36. A. Heidemann, K.J. Lushington, J.A. Morrison, K. Neumaier, and W. Press, *J. Chem. Phys.* **81**, 5799, (1984)*.

37. G.J. Vogt and K.S. Pitzer, *J. Chem. Thermodynam.* **8**, 1011 (1976).

38. I.N. Krupskii, V.M. Gasan, A.I. Prokhvatilov, *Solid State Commun.* **15**, 803 (1974).

39. A.N. Aleksandrovskii, V.B. Kokshenev, V.G. Manzhelii, and A.M. Tolkachev, *Sov. J. Low Temp. Phys.* **4**, 435 (1978), [*Fiz. Nizk. Temp.* **4**, 915 (1978)].

40. M.A. Strzhemechny and O.I. Tokar, *Sov. J. Low Temp. Phys.* **12**, 486 (1986), [*Fiz. Nizk. Temp.* **12**, 857 (1986)].

41. R.H. Beaumont, H. Chihara, and J.A. Morrison, *Proc. Phys. Soc.* (London), **78**, 1462, (1961).

42. L. Finegold and N.E. Phillips, *Phys. Rev.,* **177**, 1383, (1969).

43. M.I. Bagatskii, Thesis Doctor's, Kharkov, Ukraine, 2000.

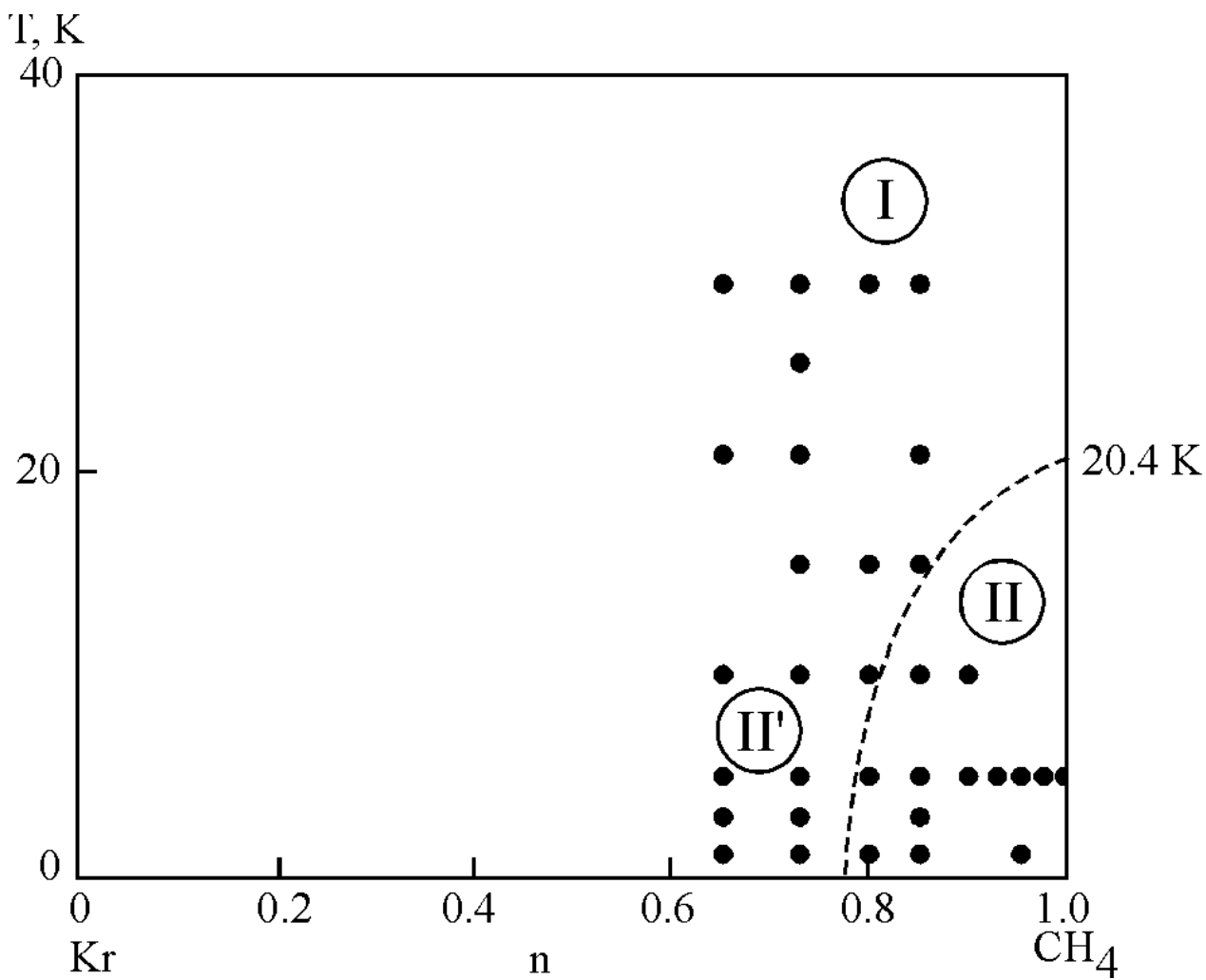

Fig. 1. The phase diagram region for solid Kr–CH$_4$ solutions (S .Grondey (1986) [3])). II – partially ordered phase, II′ – octupole glass, I – orientionally disordered phase (hindered rotation of molecules); - - - - – expected position of the interphase boundary. Points show the temperature and CH$_4$ concentration where in neutron diffraction experiments were made.

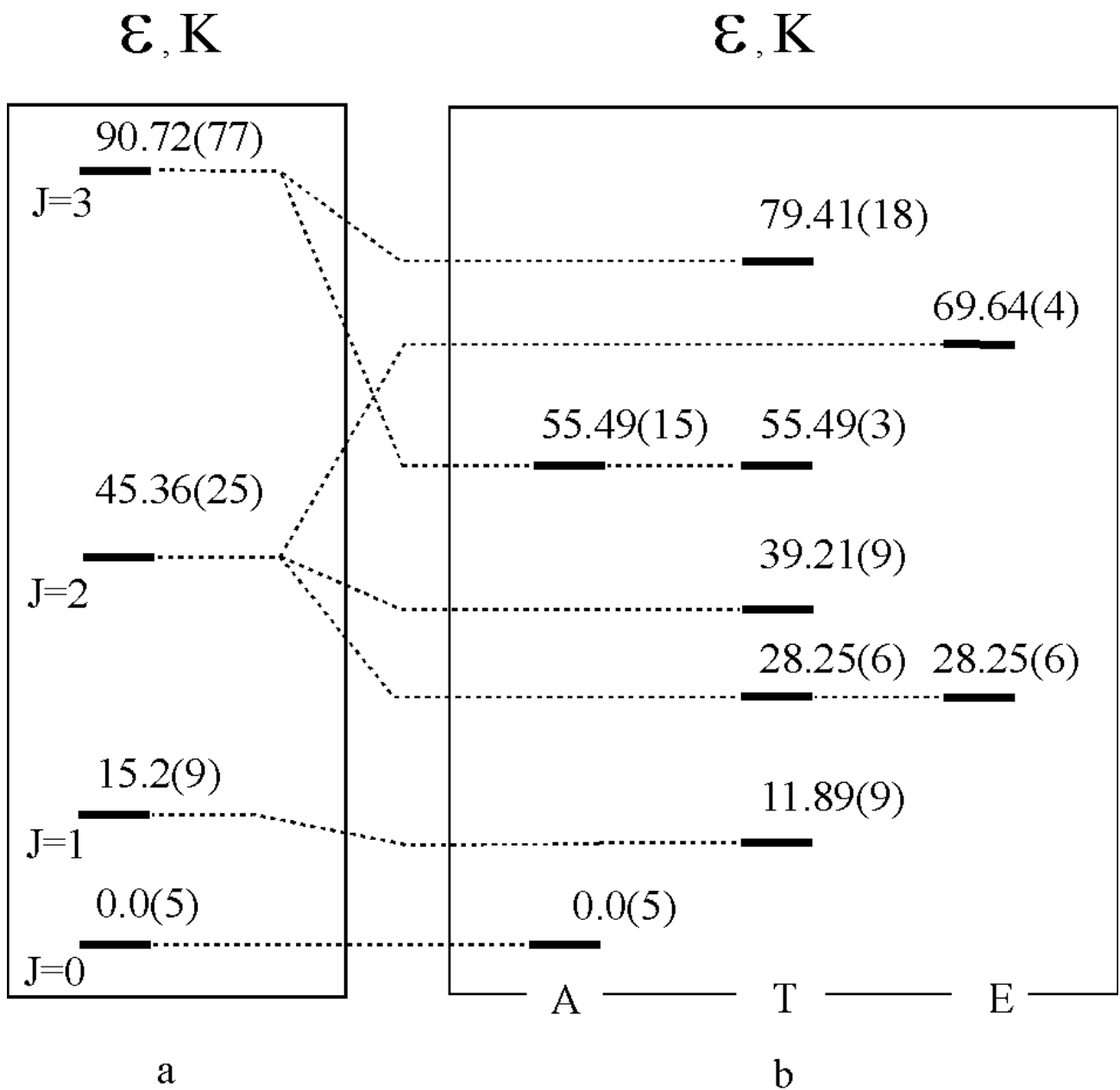

Fig. 2. Rotational energy spectrum for the A–, T–, E– modifications of CH$_4$ molecules: a – free rotator, b – solid Kr–CH$_4$ solution, calculated in [8]. $J$ – rotational quantum number; $\varepsilon$ – energy (level degeneracy's are in brackets to the right).

Table 1. The experimental $C_{sol}$ values for equilibrium vapor elasticity of solid Kr-CH$_4$ solutions with $n$ = 4.93, 9.82 [4,43]; 29.7 and 59.9 mol.% CH$_4$.

| T, K | $C_{sol}$, J/mol×K | T, K | $C_{sol}$, J/mol×K | T, K | $C_{sol}$, J/mol×K | T, K | $C_{sol}$, J/mol×K |
|---|---|---|---|---|---|---|---|
| | | | | n = 4.93 mol.% | | | |
| 1.2053 | 0.0097 | 2.043 | 0.0510 | 4.121 | 0.570 | 5.845 | 1.790 |
| 1.2703 | 0.0116 | 2.044 | 0.0515 | 4.122 | 0.562 | 6.211 | 2.14 |
| 1.4745 | 0.0178 | 2.075 | 0.0535 | 4.135 | 0.567 | 6.427 | 2.31 |
| 1.4773 | 0.0180 | 2.352 | 0.0810 | 4.573 | 0.803 | 6.427 | 2.29 |
| 1.4801 | 0.0182 | 2.434 | 0.0905 | 4.840 | 0.942 | 6.646 | 2.51 |
| 1.4804 | 0.0182 | 2.444 | 0.0930 | 4.989 | 1.095 | 7.034 | 2.88 |
| 1.4880 | 0.0186 | 2.740 | 0.134 | 4.993 | 1.075 | 7.285 | 3.11 |
| 1.6365 | 0.0251 | 2.946 | 0.177 | 5.018 | 1.080 | 7.444 | 3.25 |
| 1.6886 | 0.0274 | 3.138 | 0.227 | 5.040 | 1.100 | 7.752 | 3.65 |
| 2.036 | 0.0500 | 3.190 | 0.238 | 5.356 | 1.330 | 7.983 | 3.88 |
| 2.037 | 0.0501 | 3.465 | 0.309 | 5.409 | 1.390 | 7.987 | 3.82 |
| | | | | n = 9.82 mol.% | | | |
| 1.0414 | 0.0071 | 1.8515 | 0.0780 | 3.463 | 0.592 | 3.957 | 0.855 |
| 1.0545 | 0.0076 | 1.9316 | 0.0835 | 3.602 | 0.630 | 4.208 | 1.125 |
| 1.1251 | 0.0096 | 2.006 | 0.0935 | 3.759 | 0.816 | 4.547 | 1.330 |
| 1.1840 | 0.0128 | 2.106 | 0.122 | 3.776 | 0.831 | 4.927 | 1.685 |
| 1.2865 | 0.0169 | 2.227 | 0.152 | 3.794 | 0.748 | 5.322 | 2.10 |
| 1.3836 | 0.0254 | 2.398 | 0.200 | 3.800 | 0.776 | 5.787 | 2.56 |
| 1.4717 | 0.0316 | 2.482 | 0.229 | 3.806 | 0.804 | 6.263 | 2.93 |
| 1.5698 | 0.0397 | 2.546 | 0.224 | 3.817 | 0.733 | 6.859 | 3.50 |
| 1.6665 | 0.0505 | 3.054 | 0.370 | 3.820 | 0.817 | 6.942 | 3.58 |
| 1.7576 | 0.0620 | 3.235 | 0.470 | 3.838 | 0.797 | 7.484 | 4.12 |
| 1.8226 | 0.0790 | 3.415 | 0.595 | 3.952 | 0.905 | 7.689 | 4.22 |
| | | | | n = 29.73 mol.% | | | |
| 0.9109 | 0.0285 | 2.080 | 1.08 | 5.255 | 3.64 | 10.596 | 8.62 |
| 0.9686 | 0.0470 | 2.379 | 1.51 | 5.859 | 4.27 | 11.180 | 9.09 |
| 1.0090 | 0.0765 | 2.711 | 1.91 | 6.626 | 4.86 | 11.311 | 9.21 |
| 1.0539 | 0.0830 | 2.860 | 1.77 | 6.841 | 5.01 | 12.091 | 9.90 |
| 1.1194 | 0.120 | 2.972 | 2.01 | 7.524 | 5.57 | 12.141 | 9.91 |
| 1.2835 | 0.222 | 3.065 | 2.29 | 8.109 | 6.03 | 13.027 | 10.98 |
| 1.3610 | 0.272 | 3.217 | 2.40 | 8.753 | 6.74 | 13.965 | 11.46 |
| 1.4408 | 0.355 | 3.453 | 2.58 | 9.487 | 7.73 | 15.060 | 12.12 |
| 1.5773 | 0.498 | 3.610 | 2.64 | 9.865 | 8.08 | | |
| 1.7506 | 0.699 | 4.022 | 3.15 | 10.063 | 8.23 | | |
| 1.8760 | 0.893 | 4.928 | 3.55 | 10.445 | 8.50 | | |

| T, K | C$_{sol}$, J/mol×K | T, K | C$_{sol}$, J/mol×K | T, K | C$_{sol}$, J/mol×K | T, K | C$_{sol}$, J/mol×K |
|---|---|---|---|---|---|---|---|
| | | | | | | | |
| | | | n = 59.9 mol.% | | | | |
| 0.8629 | 0.113 | 1.6184 | 1.68 | 3.029 | 4.57 | 7.212 | 6.25 |
| 0.9156 | 0.188 | 1.7789 | 2.21 | 3.213 | 4.73 | 7.892 | 6.69 |
| 0.9509 | 0.194 | 1.8823 | 2.49 | 3.361 | 4.86 | 8.594 | 7.35 |
| 1.0240 | 0.310 | 1.9261 | 2.61 | 3.622 | 4.83 | 9.313 | 8.27 |
| 1.0775 | 0.385 | 2.000 | 2.79 | 3.726 | 4.85 | 10.097 | 9.11 |
| 1.0888 | 0.395 | 2.088 | 3.08 | 4.363 | 4.95 | 11.932 | 10.37 |
| 1.1711 | 0.510 | 2.187 | 3.33 | 4.989 | 5.04 | 13.085 | 11.52 |
| 1.2881 | 0.820 | 2.441 | 3.82 | 5.655 | 5.31 | 14.458 | 12.01 |
| 1.3618 | 1.09 | 2.582 | 4.17 | 6.118 | 5.53 | 16.090 | 13.74 |
| 1.4358 | 1.16 | 2.712 | 4.35 | 6.593 | 5.81 | 17.888 | 15.61 |
| 1.5730 | 1.50 | 2.873 | 4.51 | 7.123 | 6.18 | 19.654 | 16.02 |

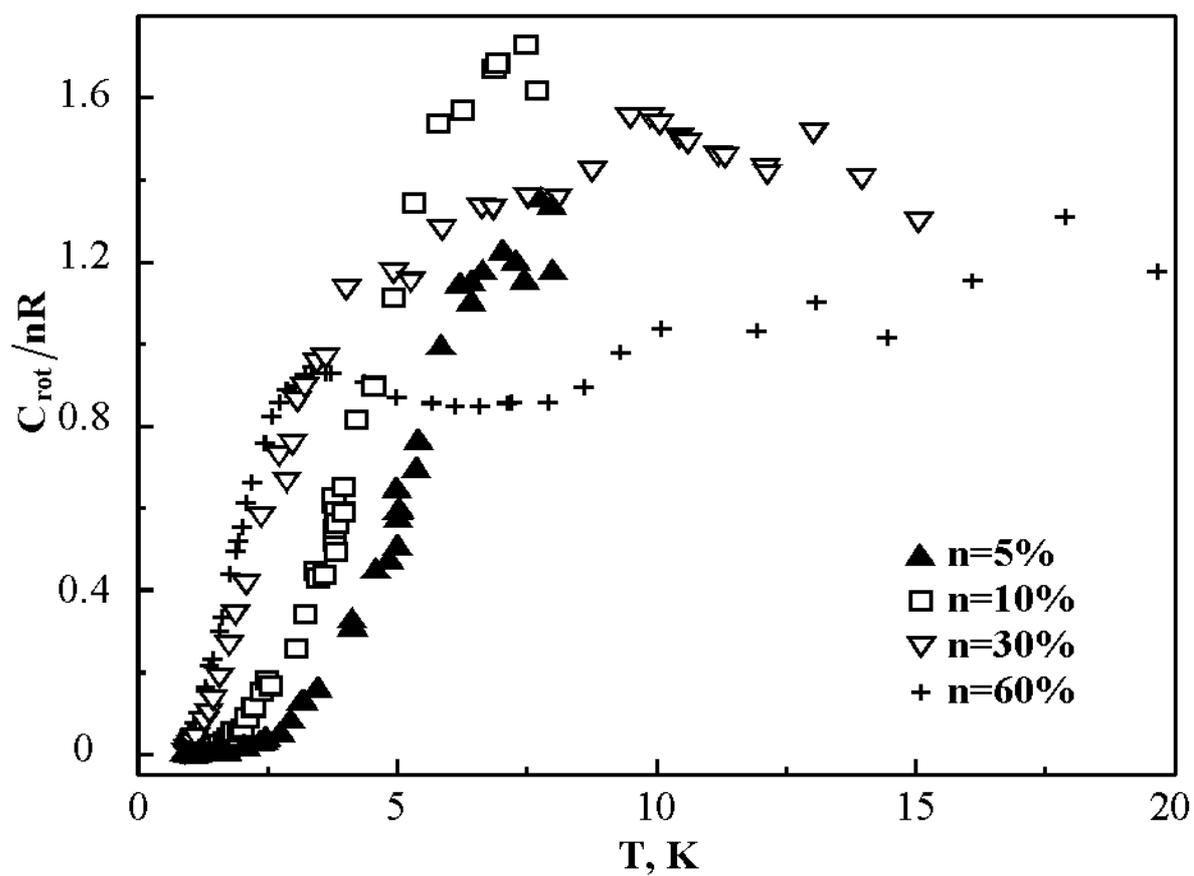

Fig. 3. Normalized heat capacity $C_{rot}/Rn$ of solid   solutions $(CH_4)_nKr_{1-n}$.

Table 2. Relative concentrations of $CH_4$ clusters and $CH_4$ molecules in the clusters in a binary solution (fcc lattice, random component distribution in the solution): $n$– $CH_4$ concentration, $n_1$ – concentration of singles ($n_1=(1-n)^{12}$); $n_2$ – concentration of two particle clusters ($n_2 = 6n(1-n)^{18}$); $(1 - n_1)$ – concentration of particles in clusters without singles; $(1 - n_1 - 2n_2)$ – concentration of particles in clusters with more than two particles.

| n | $n_1$ | $n_2$ | $1-n_1$ | $1-n_1-2n_2$ |
|---|---|---|---|---|
| 0.01 | 0.886 | $5.01 \times 10^{-2}$ | 0.114 | $1.4 \times 10^{-2}$ |
| 0.02 | 0.785 | $8.34 \times 10^{-2}$ | 0.215 | $4.8 \times 10^{-2}$ |
| 0.05 | 0.540 | 0.119 | 0.460 | 0.221 |
| 0.10 | 0.282 | $9.01 \times 10^{-2}$ | 0.718 | 0.537 |
| 0.20 | $6.87 \times 10^{-2}$ | $2.16 \times 10^{-2}$ | 0.931 | 0.888 |
| 0.30 | $1.38 \times 10^{-2}$ | $2.93 \times 10^{-3}$ | 0.986 | 0.980 |

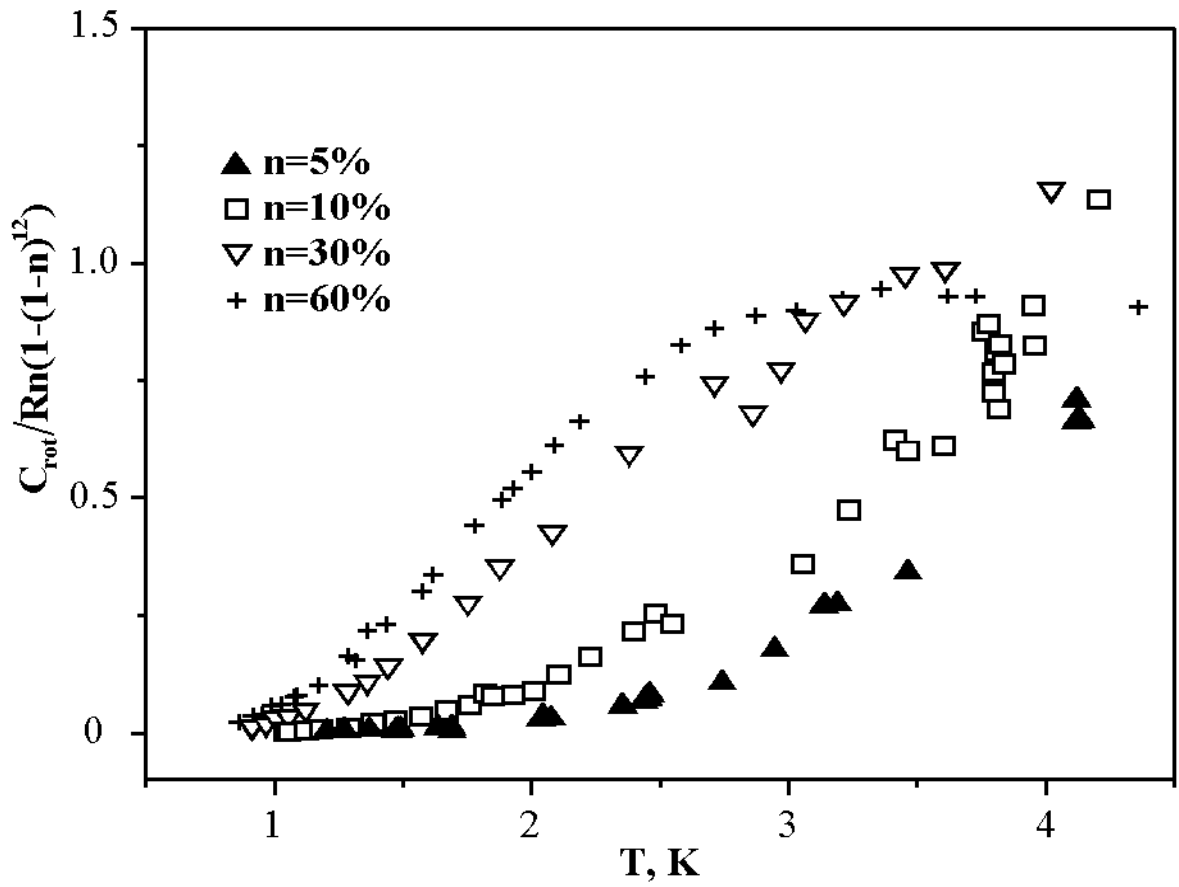

Fig. 4. Normalized heat capacity $C_{rot}/Rn(1-(1-n)^{12})$ of solid solutions $(CH_4)_nKr_{1-n}$.

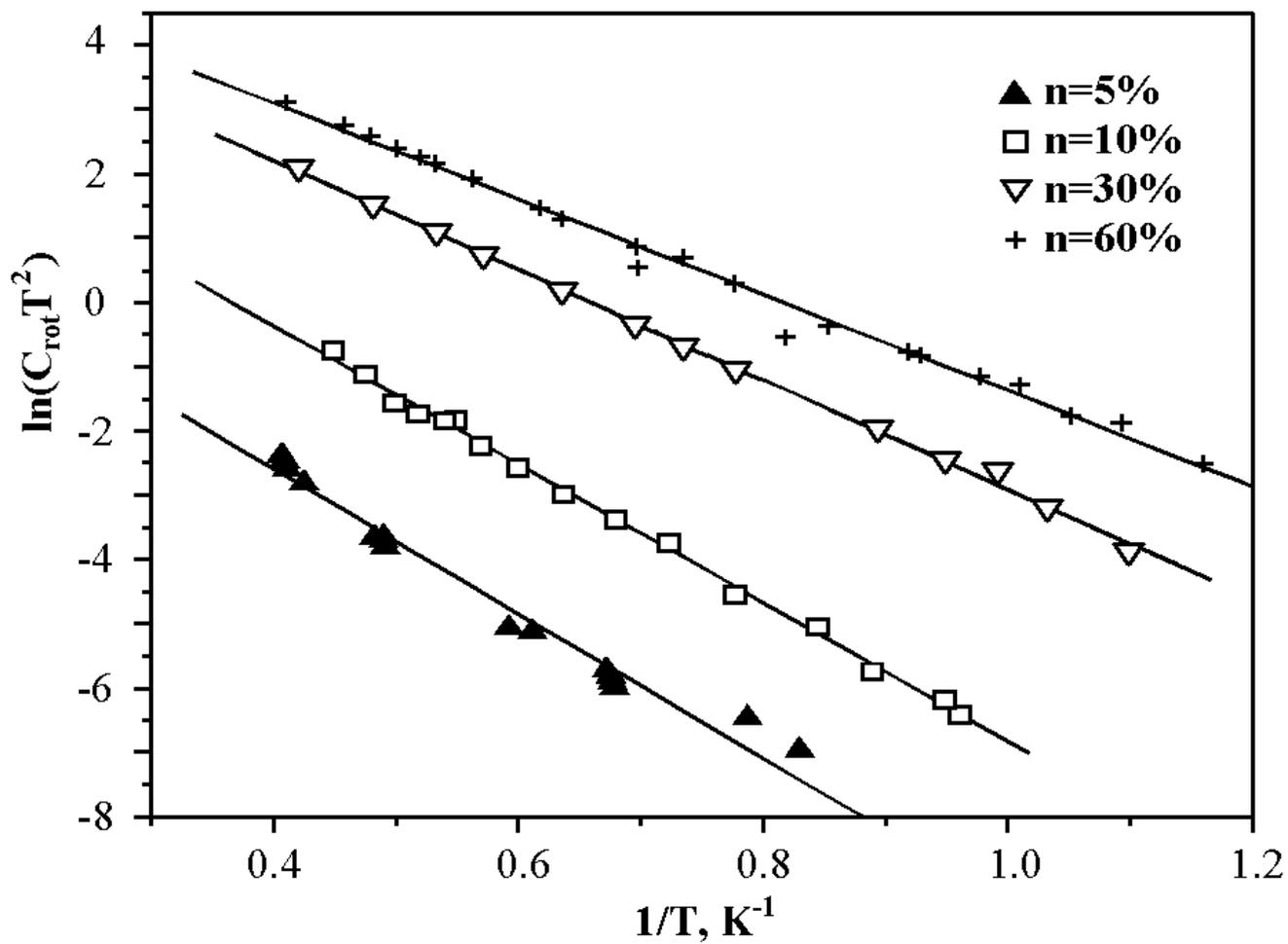

Fig. 5. Temperature dependences of heat capacity $C_{rot}$ of solid solutions $(CH_4)_n Kr_{1-n}$.

Fig. 6. Concentration dependences of mean (effective) energies $\varepsilon_{AT}$ for solid solutions $(CH_4)_n Kr_{1-n}$.

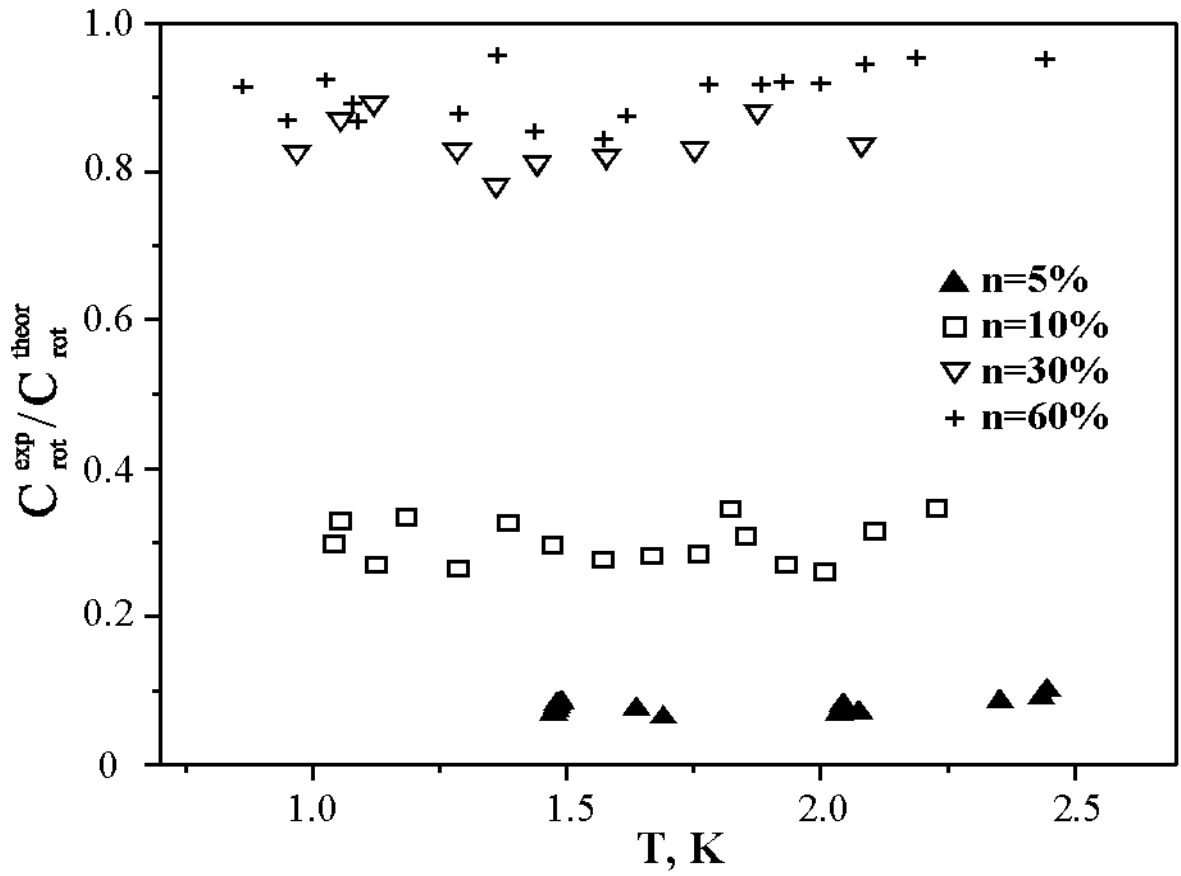

Fig. 7. Temperature dependences $C_{rot}^{exp}/C_{rot}^{theor}$ for solid solutions $(CH_4)_n Kr_{1-n}$. $C_{rot}^{theor}$ is calculated for $\varepsilon_{AT} = 11.7$ K $(n = 5\%)$; $\varepsilon_{AT} = 10.8$ K $(n = 10\%,)$; $\varepsilon_{AT} = 8.5$ K $(n = 30\%,)$; $\varepsilon_{AT} = 7.4$ K $(n = 60\%,)$.

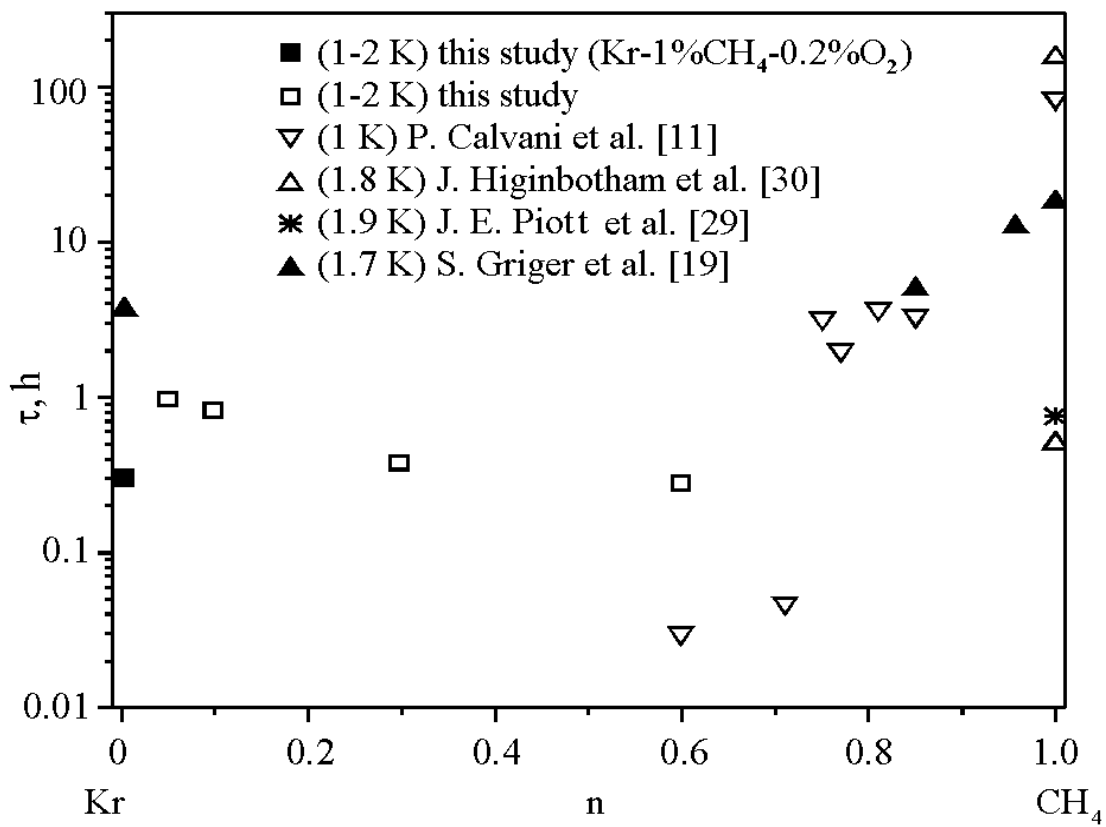

Fig. 8. Concentration dependences of characteristic mean conversion times $\tau$ in solid solutions $(CH_4)_n Kr_{1-n}$ and $Kr–1\%$ $CH_4–0.2\%$ $O_2$.

Table 3. Parameters determining heat capacity $C_{rot}^{exp}$ of the rotational subsystem in solid binary $(CH_4)_n Kr_{1-n}$ and ternary $(CH_4)_{0.01} Kr_{0.988} (O2)_{0.002}$ solutions at $T < 2.5$ K.

| n | 0.01(with $O_2$) | 0.05 | 0.10 | 0.30 | 0.60 |
|---|---|---|---|---|---|
| $\varepsilon_{AT}$ , K | 11.7±0.4 | 11.7±0.7 | 10.8±0.6 | 8.5±0.5 | 7.4±0.5 |
| $K'$, % | 72 | 8 | 30 | 84 | 91 |
| $K''$, % | - | 17 | 41 | 85 | 91 |
| $t_m$, min | 23 | 11 | 27 | 44 | 41 |
| $\tau$, min | 18 | 58 | 50 | 23 | 17 |

Nomenclature: $\varepsilon_{AT}$ – energy spacing between the lowest levels of A– and T– $CH_4$ modifications; $t_m$ – characteristic mean time of one heat capacity measurement; $\tau$ – characteristic mean time of $CH_4$ conversion; $K'$ – ratio between the number of $CH_4$ molecules converted from the $\varepsilon_{oA}$ state to the $\varepsilon_{oT}$ state in real experiment during the time $t_m$ and the corresponding number of molecules for the equilibrium distribution of the nuclear spin modifications of $CH_4$; $K''$ is similar to $K'$ for $CH_4$ molecules in clusters